# $1/f^{\gamma}$ Tunnel Current Noise through Si-bound Alkyl Monolayers

Nicolas Clément[1], Stéphane Pleutin[1], Oliver Seitz[2], Stéphane Lenfant[1] and Dominique Vuillaume[1,a]

[1] "Molecular Nanostructures and Devices" group, Institute for Electronics, Microelectronics and Nanotechnology, CNRS and University of Sciences and Technologies of Lille, BP60069, Avenue Poincaré, F-59652 cedex, Villeneuve d'Ascq, France

[2] Department of Materials and Interfaces, Weizmann Institute of Science, Rehovot, Israel

## Abstract

We report low frequency tunnel current noise characteristics of an organic monolayer tunnel junction. The measured devices, n-Si/alkyl chain ($C_{18}H_{37}$)/Al junctions, exhibit a clear $1/f^{\gamma}$ power spectrum noise with $1< \gamma <1.2$. We observe a slight bias-dependent background of the normalized current noise power spectrum ($S_I/I^2$). However, a local increase is also observed over a certain bias range, mainly if $V > 0.4$ V, with an amplitude varying from device to device. We attribute this effect to an energy-dependent trap-induced tunnel current. We find that the background noise, $S_I$, scales with $(\partial I / \partial V)^2$. A model is proposed showing qualitative agreements with our experimental data.

[a] Corresponding author : dominique.vuillaume@iemn.univ-lille1.fr

# I. INTRODUCTION

Molecular electronics is a challenging area of research in physics and chemistry. Electronic transport in molecular junctions and devices has been widely studied from a static (dc) point of view.[1,2] More recently electron – molecular vibration interactions were investigated by inelastic electron tunneling spectroscopy.[3] In terms of the dynamics of a system, fluctuations and noise are ubiquitous physical phenomena. Noise is often composed of $1/f$ noise at low frequency and shot noise at high frequency. Although some theories about shot noise in molecular systems were proposed,[4] it is only recently that it was measured, in the case of a single $D_2$ molecule.[5] Low frequency $1/f$ noise was studied in carbon nanotube transistors,[6] but, up to now, no study of the low frequency current noise in molecular junctions (e.g., electrode/short molecules/electrode) has been reported. Low frequency noise measurements in electronic devices usually can be interpreted in terms of defects and transport mechanisms.[7] While it is obvious that $1/f$ noise will be present in molecular monolayers as in almost any system, only a detailed study can lead to new insights in the transport mechanisms, defect characterization and coupling of molecules with electrodes.

We report here the observation and detailed study of the $1/f^{\gamma}$ power spectrum of current noise through organic molecular junctions. n-Si/$C_{18}H_{37}$/Al junctions were chosen for these experiments because of their very high quality, which allows reproducible and reliable measurements.[8] The noise current power spectra ($S_I$) are measured for different biases. Superimposed on the background noise, we observe noise bumps over a certain bias range and propose a model that includes trap-induced tunnel current, which satisfactorily describes the noise behaviour in our tunnel molecular junctions.

## II. CURRENT-VOLTAGE EXPERIMENTS

Si-C linked alkyl monolayers were formed on Si(111) substrates (0.05-0.2 Ω.cm) by thermally induced hydrosilylation of alkenes with Si:H, as detailed elsewhere.[8,9] 50 nm thick aluminium contact pads with different surface areas between $9\times10^{-4}$ cm$^2$ and $4\times10^{-2}$ cm$^2$ were deposited at 3 Å/s on top of the alkyl chains. The studied junction, Si-n/C$_{18}$H$_{37}$/Al, is shown in Fig.1-a (inset). Figure 1-a shows typical current density – voltage (*J-V*) curves. We measured 13 devices with different pad areas. The maximum deviation of the current density between the devices is not more than half an order of magnitude. It is interesting to notice that although devices A and C have different contact pad areas (see figure caption), their *J-V* curves almost overlap. This confirms the high quality of the monolayer.[9] Figure 1-b shows a linear behaviour around zero bias and we deduce a surface-normalized conductance of about $2\text{-}3\times10^{-7}$ S.cm$^{-2}$. For most of the measured devices, the *J-V* curves diverge from that of device C at *V* > 0.4 V, with an increase of current that can reach an order of magnitude at 1 V (device B). Taking into account the difference of work functions between n-Si and Al, considering the level of doping in the Si substrate (resistivity ~ 0.1 Ω.cm), there will be an accumulation layer in the Si at *V* > -0.1 V.[10] From capacitance-voltage (*C-V*) and conductance-frequency (*G-f*) measurements (not shown here), we confirmed this threshold value (± 0.1 V). As a consequence, for positive bias, we can neglect any large band bending in Si (no significant voltage drop in Si). The *J-V* characteristics are then calculated with the Tsu-Esaki formula[11] that can be recovered from the tunnelling Hamiltonian.[12] Assuming the monolayer to be in between two reservoirs of free quasi-electrons and the system to be invariant with respect to translation in the transverse directions (parallel to the electrode plates) we get

$$J(V) = \frac{emk_B\theta}{2\pi^2\hbar^3} \int_0^{+\infty} dE\, T(E) \ln\left(\frac{1+e^{\beta(\mu-E)}}{1+e^{\beta(\mu-eV-E)}}\right) \qquad (1)$$

where $e$ is the electron charge, $m$ the effective mass of the charge carriers within the barrier, $k_B$ the Boltzmann constant, $\hbar$ the reduced Planck constant, $\mu$ the Fermi level and $\beta = 1/k_B\Theta$ ($\Theta$ the temperature in K). $T(E)$ is the transfer coefficient for quasi-electrons flowing through the tunnel barrier with longitudinal energy $E$. The total energy, $E_T$, of quasi-electrons is decomposed into a longitudinal and a transverse component, $E_T=E+E_t$; $E_t$ was integrated out in Eq. (1). The transfer coefficient is calculated for a given barrier height, $\Phi$, and thickness, $d$, and shows two distinct parts: $T(E)=T_1(E)+T_2(E)$. $T_1(E)$ is the main contribution to $T(E)$ that describes transmission through a defect-free barrier. $T_2(E)$ contains perturbative corrections due to assisted tunnelling mechanisms induced by impurities located at or near the interfaces. The density of defects is assumed to be sufficiently low to consider the defects as independent from each other, each impurity at position $\vec{r}_i$ interacting with the incoming electrons via a strongly localized potential at energy $U_i$, $U_i\delta(\vec{r}-\vec{r}_i)$. The value of $U_i$ is random. We write $T_2(E) = \sum_{i=1}^{N_{imp}} T_2(E,U_i)$, with $N_{imp}$ being the number of impurities and $T_2(E,U_i)$ the part of the transmission coefficient due to the impurity $i$. The two contributions of $T(E)$ are calculated following the method of Appelbaum and Brinkman.[13] Using Eq. (1), we obtain a good agreement with experiments. The theoretical $J$-$V$ characteristic for device C and B are shown in Fig. 1-a. The best fits are obtained with $\Phi$ = 4.7 eV, $m$ = 0.614 $m_e$ ($m_e$ is the electron mass), $10^{10}$ traps/cm² uniformly distributed in energy for device C and additional $10^{13}$ traps/cm² for device B distributed according to a Gaussian peaked at 3 eV. The transfer coefficients $T_2(E,U_i)$ show pronounced quasi-resonances at energies depending on $U_i$ that explain the important increase of current. The thickness is kept fixed, $d = 2$ nm (measured by ellipsometry[8]).

## III. NOISE BEHAVIOR

The difference observed in the *J-V* curves are well correlated with specific behaviours observed in the low frequency noise. Figure 2 shows the low frequency current noise power spectrum $S_I$ for different bias voltages from 0.02 V to 0.9 V. All curves are almost parallel and follow a perfect $1/f^\gamma$ law with $\gamma = 1$ at low voltages, increasing up to 1.2 at 1 V. We could not observe the shot noise because the high gains necessary for the amplification of the low currents induce a cut-off frequency of our current preamplifier lower than the frequency of the $1/f$ – shot noise transition. At high currents, $1/f^\gamma$ noise was observed up to 10 kHz.

The low frequency $1/f$ current noise usually scales as $I^2$, where $I$ is the dc tunnel current,[14] as proposed for example by the standard phenomenological equation of Hooge [15] $S_I = \alpha_H I^2 / N_c f$ where $N_c$ is the number of free carriers in the sample and $\alpha_H$ is a dimensionless constant frequently found to be $2 \times 10^{-3}$. This expression was used with relative success for homogeneous bulk metals[14,15] and more recently also for carbon nanotubes.[6] Similar relations were also derived for $1/f$ noise in variable range hopping conduction.[16] In Fig. 3-a we present the normalized current noise power spectrum ($S_I/I^2$) at 10 Hz (it is customary to compare noise spectra at 10 Hz) as a function of the bias *V* for devices B and C. Device C has a basic characteristic with the points following the dashed line asymptote. We use it as a reference for comparison with our other devices.

We basically observed that $S_I/I^2$ decreases with $|V|$. For most of our samples, in addition to the background normalized noise, we observe a local (Gaussian with $V$) increase of noise at $V > 0.4$ V. The amplitude of the local increase varies from device to device. This local increase of noise is correlated with the increase of current seen in the *J-V* curves. The *J-V* characteristics (Fig. 1-a) of device B diverge from those of device C at $V > 0.4$ V and this is consistent with the local increase of noise observed in Fig. 3. The observed excess noise bump is likely attributed to this Gaussian distribution of traps centred at 3 eV responsible for the current increase. Although the microscopic mechanisms associated with conductance fluctuations are not clearly identified, it is believed that the underlying mechanism involves the trapping of charge carriers in localized states.[17]. The nature and origin of these traps is however not known. We can hypothesis that the low density of traps uniformly distributed in energy may be due to Si-alkyl interface defects or traps in the monolayer, while the high density, peaked in energy, may be due to metal-induced gap states (MIGS)[18] or residual aluminum oxide at the metal-alkyl interface. The difference in the noise behaviours of samples B and C simply results from inhomogeneities of the metal deposition, i.e. of the chemical reactivity between the metal and the monolayer, or is due to the formation of a residual aluminum oxide due to the presence of residual oxygen in the evaporation chamber. More 1/f noise experiments on samples with various physical and chemical natures of the interfaces are in progress to figure out how the noise behaviour depends on specific conditions such as the sample geometry, the metal or monolayer quality, the method used for the metal deposition and so forth.

# IV. TUNNEL CURRENT NOISE MODEL

To model the tunnel current noise in the monolayers, we assume that some of the impurities may trap charge carriers. Since we do not know the microscopic details of the trapping mechanisms and the exact nature of these defects, we use a qualitative description that associates to each of them an effective Two-Level Tunnelling Systems (TLTS) characterized by an asymmetric double well potential with the two minima separated in energy by $2\varepsilon_i$. We denote as $\Delta_i$ the term allowing tunneling from one well to the other, and get, after diagonalization, two levels that are separated in energy by $E_i = \sqrt{\varepsilon_i^2 + \Delta_i^2}$. Since we are interested in low frequency noise, we focus on defects with very long trapping times i.e. defects for which $\Delta_i \ll \varepsilon_i$. The lower state (with energy $E_i^-$) corresponds to an empty trap, the upper state (with energy $E_i^+$) to a charged one. The relaxation rate from the upper to the lower state is determined by the coupling with the phonons and/or with the quasi-electrons giving $\tau^{-1} \propto E_i \Delta_i^2 \coth\dfrac{E_i}{2k_B\theta}$ and $\tau^{-1} \propto \dfrac{\Delta_i^2}{\hbar E_i} \coth\dfrac{E_i}{2k_B\theta}$, respectively. In all cases, the time scale of the relaxation, $\tau$, is very long compared to the duration of a scattering event. This allows us to consider the TLTS with a definite value at any instant of time. We then consider the following spectral density of noise for each TLTS[19]

$$S_I^i(f) = \overline{(I_- - I_+)^2} \dfrac{\tau}{1+\varpi^2\tau^2} \operatorname{Cosh}^{-2}\dfrac{E_i}{2k_B\theta} \tag{2}$$

where $\varpi = 2\pi f$ and $I_{-(+)}$ is the tunnel current for the empty (charged) impurity state. In this equation, we consider the average of $(I_- - I_+)$ over the TLTSs, having similar $\varepsilon_i$ and $\Delta_i$. The difference between the two levels of current has two different origins. The first one is the change in energy of the impurity level that directly affects $T_2(E)$. The second one is the change in the charge density at the interfaces of the molecular junction induced by the trapped quasi-electron that produces a shift in the applied bias, $\delta V$. We write

$$I_+(V) \cong I_-(V+\delta V) + \mathsf{A}\frac{emk_B\theta}{2\pi^2\hbar^3}\int_0^{+\infty}dE\left.\frac{\partial T_2(E,U_i)}{\partial U_i}\right|_{E_i^-} E_i \ln\left(\frac{1+e^{\beta(\mu-E)}}{1+e^{\beta(\mu-eV-\delta V-E)}}\right) \quad (3)$$

where $\mathsf{A}$ is the junction (metal electrode) area. The first term in the right hand side is due to the fluctuating applied bias, the second to the change in the impurity energy. Since $T_2(E)$ is already a perturbation, the second contribution is in general negligible but becomes important to explain the excess noise. We focus first on the background noise and therefore we keep only the first term of Eq. (3). We assume for simplicity that all the charged impurities give the same shift of bias $\delta V = e/C_{TJ}\mathsf{A}$, where $C_{TJ}$ is the capacitance of the tunnel junction per unit surface. Capacitance-voltage measurements (not shown) indicate that $C_{TJ}$ is constant for positive bias. By using usual approximations regarding the distribution in relaxation times, $\tau$, and energies, $E_i$,[19] we get

$$S_I \propto E^* \frac{1}{\mathsf{A}} N_{imp}^* \left(\left.\frac{\partial I}{\partial V}\right|_-\right)^2 \frac{e^2}{C_{TJ}^2}\frac{1}{f}. \quad (4)$$

We assume that the distribution function of $\varepsilon_i$ and $\Delta_i$, $P(\varepsilon_i,\Delta_i)$ is uniform to get the $1/f$ dependence. In this last expression, the derivative of the current is evaluated for the lower impurity state, $N_{imp}^*$ is the impurity density per unit energy and surface area. We have $E^* = E_{max}$, the maximum of $E_i$, if $E_{max} \ll k_B\theta$ and $E^* = k_B\theta$ if $E_{max} \gg k_B\theta$.

$N^*_{imp}$ cannot be determined accurately from the last equation because of lack of information concerning the microscopic nature of the traps.

This predicted dependence of $S_I$ on $(\partial I/\partial V)^2$ is experimentally verified in Fig.4, where $S_I$ vs. $(\partial I/\partial V)$ for device C is plotted on a log-log scale, showing a slope of 2. At the same time (see inset), we show that $S_I$ - $I$ follows a power law with a slope of 1.7 and not 2 as usually assumed. The value 1.7 explains why the normalized noise $S_I/I^2$ shown in Fig. 3 decreases with $V$. The appropriate normalization factor to obtain flat background noise is $S_I/I^{1.7}$. These two features imply that $(\partial I/\partial V)^2$ scales with $I^{1.7}$ which has been experimentally verified from the $I$-$V$ curves (not shown). The calculated noise, using Eq. (4), is shown in Fig. 3.b. Qualitative agreements with the experimental data are obtained. With few defects uniformly distributed (device C), $S_I/I^2$ follows (green solid line) at low voltages the dashed line asymptote. With additional defects with a Gaussian distribution (device B), a local increase is found at the correct position but with much too small amplitude (dot blue line). To get a better estimate it is essential to take into account the second term of Eq. (3). Results are shown in Fig. 3.b (blue solid line) taking $E^*=5e\delta V$. The quasi resonances of $T_2(E)$ are at the origin of the local increase. The Gaussian distribution selects defects for which $T_2(E,U_i)$ shows quasi resonance in the appropriate range of energy. These traps may be associated to a non-uniform contribution to the distribution function $P(\varepsilon_i,\Delta_i)$ that would break the $1/f$ dependence of $S_I$ above certain bias. This is what is observed in Fig. 2, with $\gamma$ changing from 1 to 1.2.

## V. CONCLUSION

In summary, we have reported the study of low frequency ($1/f^{\gamma}$) current noise in molecular junctions. We have correlated the small dispersion observed in dc *J-V* characteristics and the local increase of normalized noise at certain biases (mainly at *V* > 0.4 V). A theoretical model qualitatively explains this effect as due to the presence of an energy-localized distribution of traps. The model predicts that the power spectrum of the background current noise is proportional to $(\partial I/\partial V)^2$ as observed in our experiments.[20] We also show that the power spectrum of the current noise should be normalized as $S_I/I^{1.7}$. The background noise is associated with a low density of traps uniformly distributed in energy that may be due to Si-alkyl interface defects or traps in the monolayer. The local increase of noise for bias *V*>0.4V is ascribed to a high density of traps, peaked in energy, probably induced by the metal deposition on the monolayer.

## ACKNOWLEDGEMENTS

We thank David Cahen for many valuable discussions. N.C. and S.P. acknowledge support from the "ACI nanosciences" program and IRCICA. We thank Hiroshi Inokawa, Frederic Martinez for helpful comments.

**Figure Captions:**

**Fig.1:** **(a)** Experimental *J-V* curves at room temperature for n-Si/$C_{18}H_{37}$/Al junctions. The contact areas are 0.36 mm$^2$ for device A and 1 mm$^2$ for devices B and C. The voltage *V* is applied to the aluminium pad and the Si is grounded, using a semiconductor signal analyzer Agilent 4155C. Each curve was acquired with a trace-retrace protocol and repeated 3 times with different delay times between each measurement (voltage step *ΔV*=1 mV) in order to check a possible hysteresis effect and confirm that no transient affects the dc current characteristics. Theoretical *J-V* curves are also shown for devices B and C. **(b)** *J-V* curves around zero bias in a linear scale for the three samples showing the good linearity at low bias.

**Fig.2:** Low frequency ($1/f^{\gamma}$) power spectrum current noise for device C. Although we measured all spectra of the sequence |*V*|=[0.02; 0.05; 0.1; 0.15; 0.2; 0.25; 0.3; 0.35; 0.4; 0.45; 0.5; 0.6; 0.7; 0.8; 0.9; 1 V], we selected for this figure only spectra with *V* > 0 V and a spacing of 0.2 V for clearer presentation. *γ* varies from 1 at low voltages to 1.2 at 1 V. For noise measurements, the experimental setup was composed of a low noise current-voltage preamplifier (Stanford SR570), powered by batteries, and a spectrum analyser (Agilent 35670A). All the measurements were performed under controlled atmosphere ($N_2$) at room temperature.

**Fig.3: (A)** Normalized power spectrum current noise $S_I/I^2$ as a function of bias *V* for devices B and C. The curve for device C follows asymptotes (black dashed lines) which are used as a reference for other devices. A local increase of noise over the asymptotes

with a Gaussian shap (solid lines) is shown. **(B)** Theoretical estimates are shown for $V > 0$, based on Eq. (4) with a uniform defect distribution (dashed line), with adding a Gaussian energy-localized distribution of defects (thin solid line), and keeping the two terms of Eq. (3) with $E^*=5e\delta V$ (bold solid line). An ad-hoc multiplicative factor has been applied to the theoretical results.

**Fig.4:** $S_I$ - $(\partial I / \partial V)$ curve for device C on a log-log scale. The dashed line represents the slope of 2. In the inset, the $S_I$ - $I$ curve is also presented on a log-log scale with a slope of 1.7.

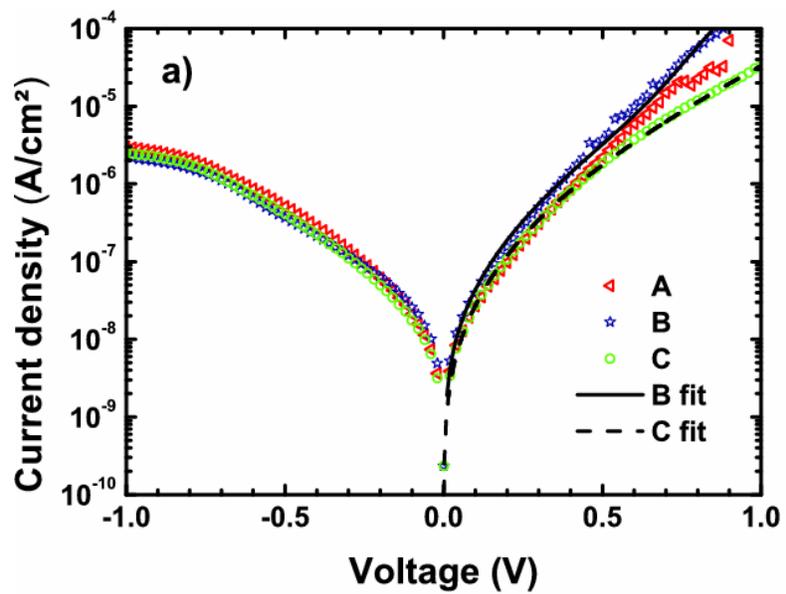

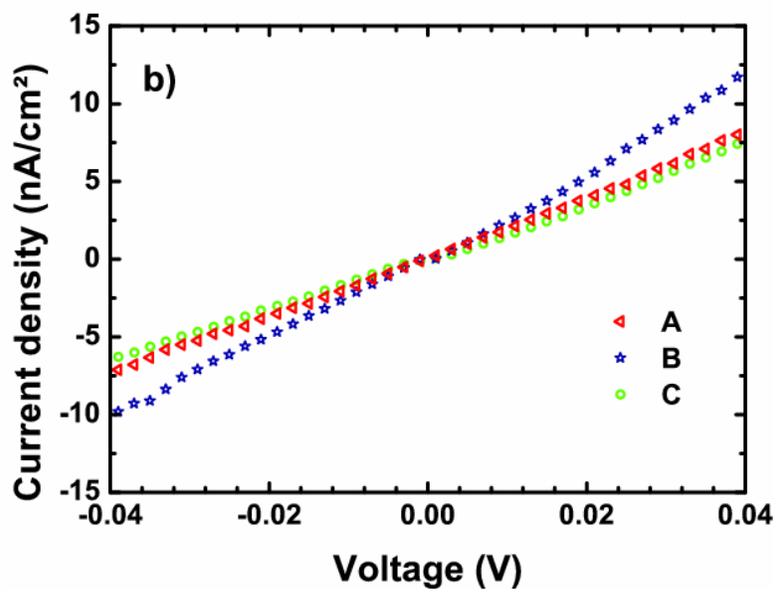

Fig.1

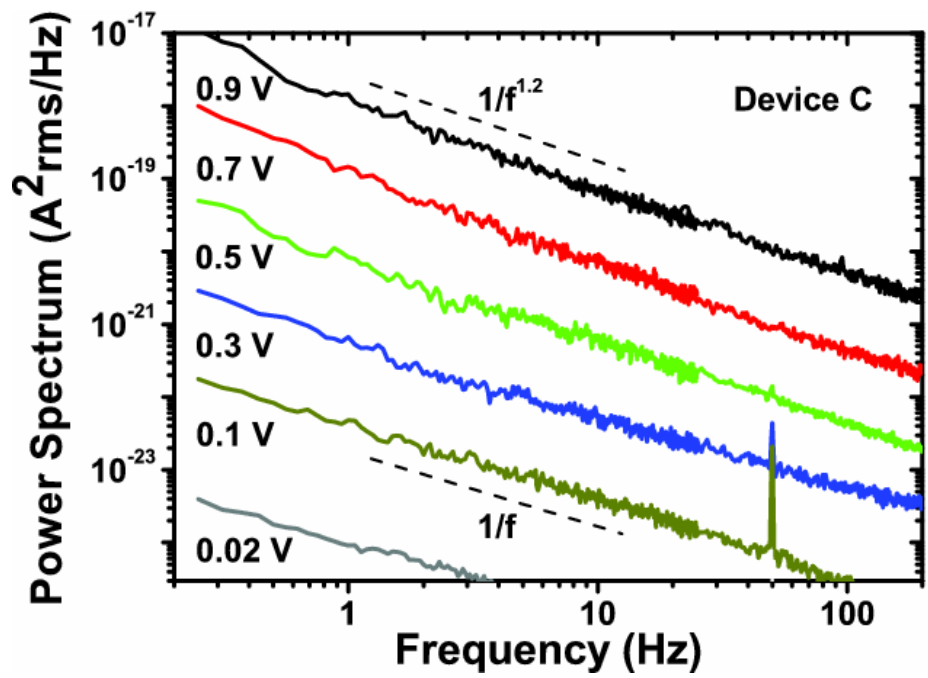

Fig.2

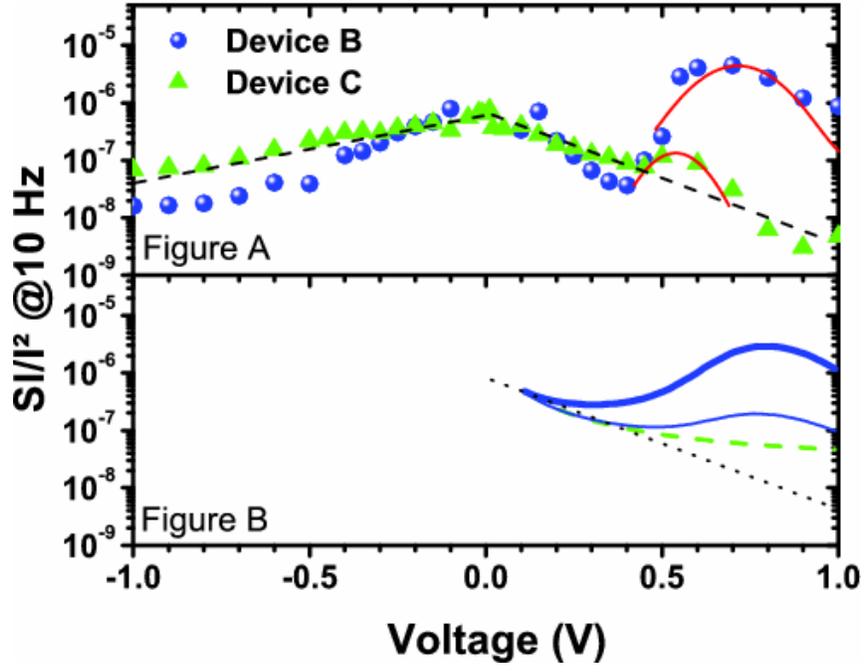

Fig.3

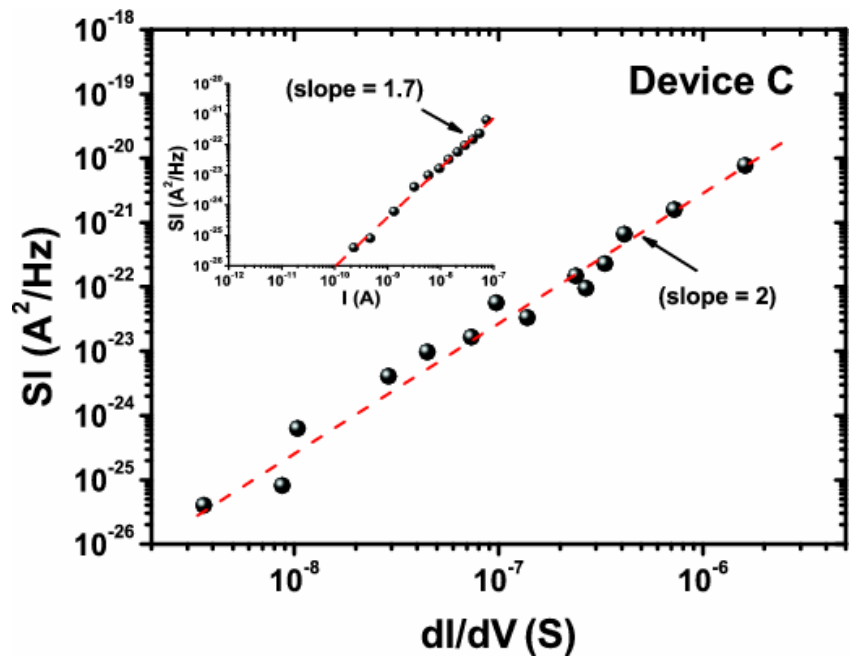

Fig.4